\documentclass[AMA,Times2COL]{WileyNJDv5}

\articletype{Article Type}%

\received{Date Month Year}
\revised{Date Month Year}
\accepted{Date Month Year}
\journal{Journal}
\volume{00}
\copyyear{2023}
\startpage{1}

\begin{document}

\title{Nonlinear Elasticity at the Damage Threshold of Semiconductor Nanocrystals}

\author[1]{Daniel Hensel}

\author[2]{Adriana Rodrigues}

\author[2]{Anagha Kamath}

\author[3]{Daniel Schmidt}

\author[1]{Mariana Brede}

\author[4]{Oliver Skibitzki}

\author[2]{Fariba Hatami}

\author[1,3]{Peter Gaal}

\authormark{Hensel \textsc{et al.}}
\titlemark{Nonlinear Elasticity at the Damage Threshold of Semiconductor Nanocrystals}

\address[1]{\orgname{Leibniz-Institut f{\"u}r Kristallz{\"u}chtung}, \orgaddress{\state{Berlin}, \country{Germany}}}

\address[2]{\orgname{Humboldt University}, \orgaddress{\state{Berlin}, \country{Germany}}}

\address[3]{\orgname{TXproducts UG}, \orgaddress{\state{Hamburg}, \country{Germany}}}

\address[4]{\orgname{IHP-Leibniz-Institut für Innovative Mikroelektronik}, \orgaddress{\state{Frankfurt (Oder)}, \country{Germany}}}

\corres{Corresponding author Peter Gaal, \email{peter.gaal@ikz-berlin.de}}



\abstract[Abstract]{The nonlinear photoacoustic response of indium phosphide nanocrystals on silicon nanotip arrays is investigated using time-resolved optical pump-probe spectroscopy and synchrotron-based X-ray diffraction. Femtosecond laser excitation triggers low-frequency and high-frequency radial breathing modes of the nanocrystals at 8 GHz and 10.3 GHz, respectively. At excitation fluences above 3\,mJ/cm$^2$, nonlinear frequency mixing occurs, including sum- and difference-frequency generation, indicative of strain-induced nonlinear elasticity. A higher-order extension of Hooke's law models the fluence-dependent spectral response and yields a physically valid elastic energy potential. Ex-situ energy-dispersive X-ray spectroscopy reveals a correlation between nanocrystal oxidation and the emergence of nonlinear acoustic modes. Time-resolved X-ray diffraction confirms the nanocrystals as the origin of the low-frequency modes and supports the hypothesis of acoustic decoupling from the substrate. These findings provide insight into the mechanical limits of semiconductor nanostructures under intense optical excitation and suggest new pathways for material characterization and optomechanical control at the nanoscale. The results advance the understanding of nonlinear phonon dynamics in nanocrystals and highlight their potential for integration into next-generation photonic and quantum devices.}


\maketitle



\section{Introduction}\label{sec:Intro}
Semiconductor (SC) nanostructures exhibit unique physical and chemical properties arising from quantum confinement and their high surface-to-volume ratios. These characteristics enable the tailoring of optical properties through tunable bandgaps, which are exploited in a wide range of optoelectronic applications, including light-emitting diodes, laser diodes, photodetectors and solar cells~\cite{Yan2025a,Shan2018a,Zhuang2019a,Teng2018,Du2019,Zhang2019b,Xu2014,Li2018}. Additionally, charge confinement and channel formation with reduced electron scattering contribute to the development of high-speed electronic and optoelectronic devices for modern communication systems~\cite{Seo2021,Besteiro2019}.
Beyond optoelectronics, SC nanostructures play a pivotal role in emerging technologies such as quantum computing~\cite{Chatterjee2021} and cryptography~\cite{Vajner2022}, biomedical imaging~\cite{Jiao2022}, catalysis~\cite{Ramos-Fernandez2025} and highly sensitive sensors~\cite{Hu2025}.
Recent experimental efforts have focused on using time-dependent acoustic strain to dynamically modulate physical parameters in SC nanostructures, enabling novel functionalities~\cite{Karzel2025,Lima2005}. One common approach involves generating strain via the inverse piezoelectric effect, where an applied electric field induces mechanical deformation. By engineering the geometry of the electrodes, specific acoustic modes can be selectively excited~\cite{Zhang2022,Datta1986}. Early pioneering studies demonstrated charge transport in acoustic strain fields~\cite{Cout2009} and modulation of photoluminescence emission~\cite{Lazic2017,Weiss2014,Fuhrmann2011} driven by acoustic waves.
An alternative method for generating acoustic strain is photoacoustics~\cite{Ruel2015,Matt2023a}. In this approach, absorption of an ultrashort laser pulse induces acoustic shock waves that propagate through the material. The frequency and mode spectrum of these waves can be precisely controlled through spatial and temporal patterning of the excitation~\cite{bohj2013a}. Advanced photoacoustic techniques have recently emerged, offering the advantage of confining excitation to the nanoscale dimensions of SC structures. With straightforward setups, it is possible to generate extremely high acoustic frequencies reaching hundreds of gigahertz or even into the terahertz range~\cite{herz2012c}. The combination of high-frequency photoacoustic excitation with coherent strain control opens new avenues for enhanced device functionality~\cite{pude2019a}.

In this work, we investigate the response of SC nanostructures to photoacoustic excitation. Both theoretical and experimental studies suggest that the elastic properties of nanoscale materials can differ significantly from their bulk counterparts. These differences influence strain-induced processes such as band structure modulation and catalytic activity and may even give rise to nonlinear acoustic interactions under high strain conditions.
We present time-resolved all-optical and X-ray diffraction (XRD) measurements of the optically induced nanomechanical response of indium phosphide (InP) nanostructures. At high excitation fluences, we observe a nonlinear acoustic response, manifested through sum- and difference-frequency mixing of the acoustic signal. We present a phenomenological model that describes the fluence-dependent structural response using a nonlinear extension to Hooke's law. Ex-situ energy-dispersive X-ray spectroscopy (EDX) measurements reveal a correlation of the nonlinear response to the oxidation of the nanocrystals (NC) at fluences close to the damage threshold of the material. 

\section{Results and Discussion}\label{sec:R&D}
\begin{figure}[htbp]
	\centering
	\includegraphics[width = \columnwidth]{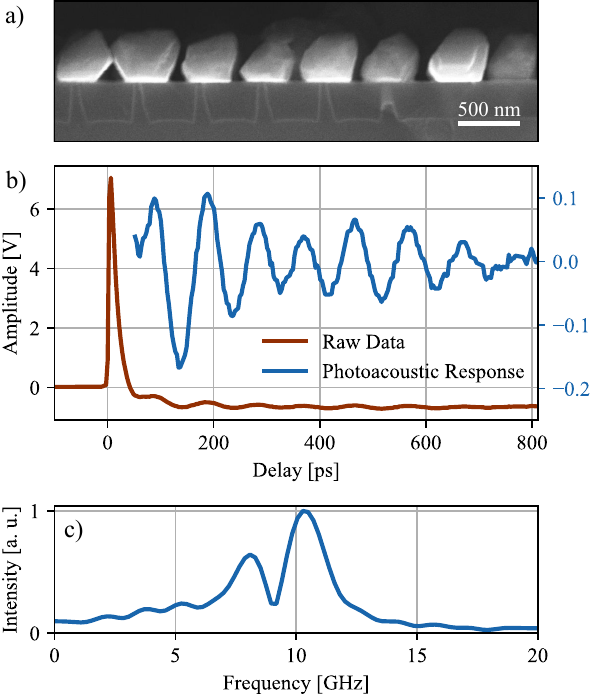}
	\caption{\textbf{(a)} SEM cross section of \textit{Sample 1}. \textbf{(b)} Transient change of the optical reflectivity after excitation with 1\,mJ/cm$^2$ (red curve) measured on the same sample structure. The blue curve depicts the photoacoustic response, i.e. the NC breathing upon the excitation. \textbf{(c)} Spectrum of the oscillation signal shown in (b) displaying the low-frequency fundamental mode at 8\,GHz and the high-frequency fundamental mode at 10.3\,GHz, respectively.}
	\label{fig:PAresponse}
\end{figure}

We investigated regular arrays of epitaxially grown InP NCs on silicon (Si) (001) nanotip substrates. A selective growth of InP on top of the nanotips is ensured by an amorphous silicon dioxide (SiO$_2$) layer that leaves only the top part of the nanotips exposed to the gaseous precursors. For a more detailed description of the sample fabrication we refer the reader to the Experimental Methods (Section~\ref{sec:Exp&Meth}). In this work we used three samples with different crystal size and spacing between the crystals. The irregular shape of the NCs can be seen in Figure~\ref{fig:PAresponse}~a), which depicts a cross-sectional scanning electron microscopy (SEM) image of \textit{Sample 1}. We measured the area of the NCs from top-view SEM images and calculated an equivalent diameter assuming a circular shape. The height was measured from the lowest to the highest point of a NC. The average dimensions of a single NC for each sample are listed in Table~\ref{tab:SampleParameters}. Note that the fabrication process guarantees only a finite precision of the particle size and pitch. Especially on \textit{Sample 1}, where the NC diameter is close to the pitch size, some NCs connect with their adjacent NC. This introduces an uncertainty in the measurement of the size and therefore the vibration eigenfrequency.

\begin{table}
    \centering
    \begin{tabular}{|c|c|c|c|}
     \hline
        & \textbf{Sample 1} & \textbf{Sample 2} & \textbf{Sample 3}\\
     \hline
        Height [nm] & $399\pm27$  & $310\pm26$ & $250\pm29$ \\
    \hline
        Diameter [nm] & $479\pm146$  &  $374\pm51$ & $288\pm33$ \\
    \hline
        Pitch [nm] & 500  &  800 & 500 \\
    \hline
        Height/Diameter & 0.83 & 0.83 & 0.87 \\
    \hline
        $f_H$ measured (calc.) [GHz] & 10.3 (8.2) & 10.2 (10.5) & 15 (13.1) \\
    \hline
        $f_L$ measured (calc.) [GHz] & 8 (6.8) & 7.3 (8.7) & 10.6 (11.4) \\
    \hline
    \end{tabular}
    \caption{Dimensions and frequencies of the low- and high-frequency modes of the average InP NC for all samples used in this study. The equivalent diameter was calculated from the average area of one NC in top-view SEM images.}
    \label{tab:SampleParameters}
\end{table}

An exemplary all-optical pump-probe measurement on \textit{Sample 1} is depicted by the red curve in Figure~\ref{fig:PAresponse}~b). We recognize a sub-picosecond reflectivity change due to hot electrons and a thermalization due to electron-phonon coupling~\cite{Shah1999a}. These features are followed by oscillations of the reflectivity amplitude (blue curve) due to the propagation of optically generated coherent acoustic phonons~\cite{Thom1986,bohj2013a,Maer2015,Ruel2015}. The spectrum of this signal is shown in Figure~\ref{fig:PAresponse}~c) and reveals two modes at 8\,GHz and 10.3\,GHz, respectively. 
We employ an analytical model of an elastic sphere~\cite{Lamb1881} to estimate the eigenfrequencies of our NCs. The model is outlined in the methods section~\ref{sec:Exp&Meth}. With a height-to-diameter ratio < 1 [see Table~\ref{tab:SampleParameters}] we expect two distinct eigenmodes with separate frequencies as depicted in Figure~\ref{fig:PAresponse}~c). The superposition of these modes leads to a beating in the recorded signal. We refer to these modes as the fundamental modes. We used literature values for InP~\cite{NICHOLS1980667} in our calculations. Note that we could determine the geometrical parameters of the NC with only limited precision. This translates directly into a deviation of the measured and predicted eigenfrequencies listed in Table~\ref{tab:SampleParameters}.

\subsection{Nonlinear Photoacoustic Response}\label{sec:PhotoacosuticExcitation}

\begin{figure*}[htbp]
	\centering
	\includegraphics[width = \textwidth]{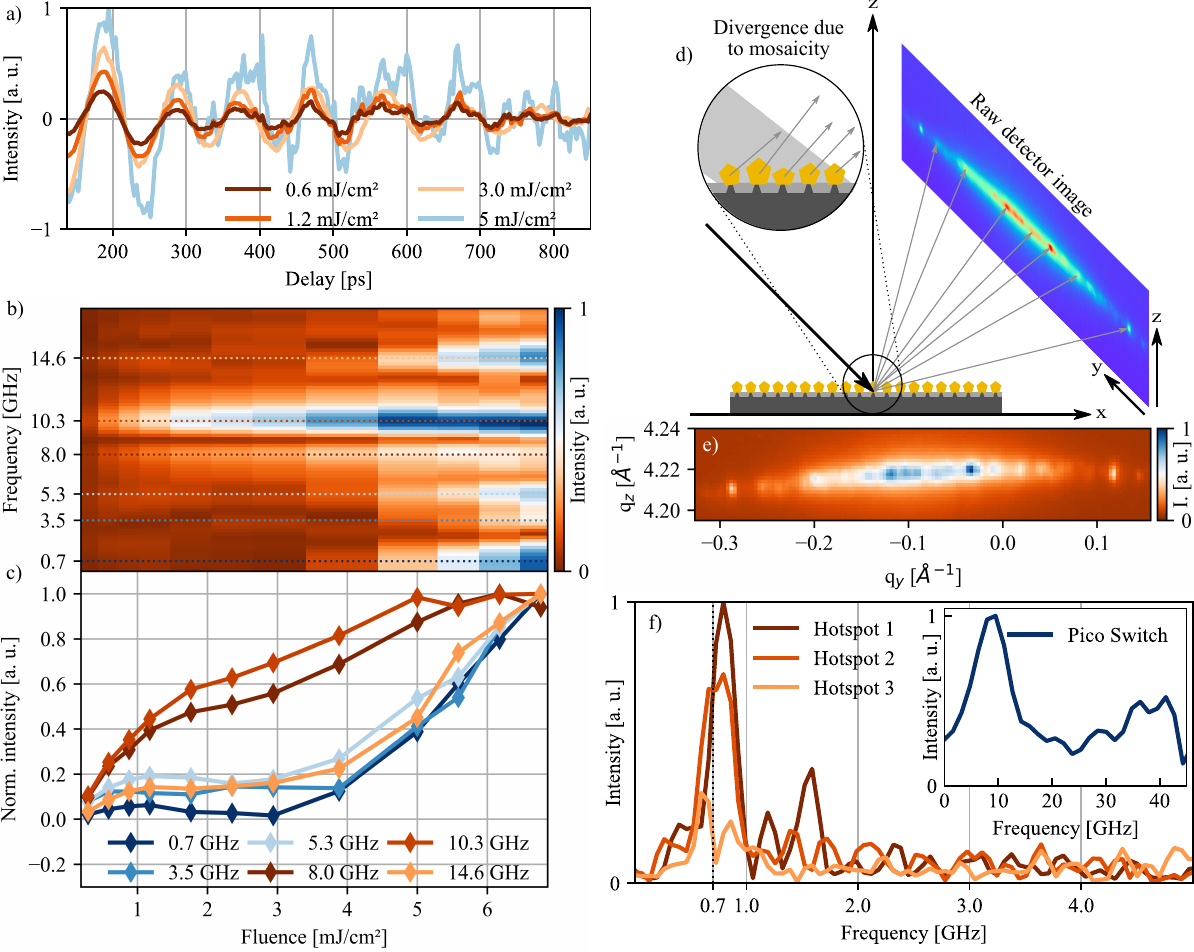}
	\caption{\textbf{(a)} Transient reflectivity from NCs measured at different optical excitation fluences on \textit{Sample 1}. We isolated the photoacoustic [c.f. Fig.~\ref{fig:PAresponse}~b)] from the electronic background. \textbf{(b)} Spectrum of the NC breathing at different optical fluences. At a fluence above 3\,mJ/cm$^{2}$, new frequency components appear in addition to the fundamental breathing modes. \textbf{(c)} Fluence dependence of the fundamental modes (red) and the nonlinear mixing products (orange and blue). \textbf{(d)} XRD measurement of the ensemble of NCs. The orientation disorder of the NC leads to a smearing of the diffracted beam on the area detector. Bright spots in the streak result from sub-ensembles of NCs with good orientation alignment. \textbf{(e)} Diffracted intensity from \textit{Sample 2} mapped into reciprocal space using the detector image depicted in in (d). \textbf{(f)} Frequency spectra of the temporal evolution of hotspots from time-resolved XRD measurements on \textit{Sample 2} with 100\,ps (red and orange) and with 10\,ps (blue) temporal resolution analyzed with a Blackman window function.}
	\label{fig:NLelasticity}
\end{figure*}

Having discussed the linear sample response to the optical excitation with a femtosecond light pulse, we now focus on the fluence dependence of the transient signal. These measurements have been again performed on \textit{Sample 1}. Figure~\ref{fig:NLelasticity}~a) depicts a series of time-resolved reflectivity measurements recorded at increasing excitation fluence. For a weak excitation (red and orange curves), we reproduce the result already shown in Figure~\ref{fig:PAresponse}~b). With increasing fluence (light blue curve) the transient response changes in amplitude and temporal shape. To better analyze the fluence dependence, we transform the measured data to the frequency domain. Figure~\ref{fig:NLelasticity}~b) depicts the spectral response up to a frequency of 18\,GHz for increasing fluence from 0.3\,mJ/cm$^{2}$ to 6.8\,mJ/cm$^{2}$. At a low excitation fluence, we obtain similar results as already depicted in Figure~\ref{fig:PAresponse}~c). For fluences higher than 3\,mJ/cm$^{2}$, we observe frequency components higher and lower than the fundamental eigenmodes of the structure. We denote these signals the non-fundamental modes. We highlight this behavior by plotting the intensity at selected frequency components in Figure~\ref{fig:NLelasticity}~c). 

Due to the limited delay range of the optical pump-probe setup of 850\,ps, we resolve the low-frequency oscillation with limited accuracy. We therefore performed time-resolved X-ray diffraction measurements at beamline ID09 at the European Synchrotron ESRF~\cite{camm2009}. These measurements were performed on \textit{Sample 2} from Table~\ref{tab:SampleParameters}. Probing the sample response with XRD at a synchrotron enables us to measure the NC vibration over a long delay range, thus resolving even low-frequency deformation modes. In addition, we can also pinpoint the InP NCs as the origin of the oscillation via the symmetric (004) Bragg reflex, which is distinct from the Si substrate or the SiO$_{2}$ layer. Reciprocal space maps (RSM) measured at a laboratory source and at beamline ID09 are provided in Figure S1 in the supplementary information. We employ a footprint of the X-ray probe beam with a diameter of 30\,$\mu$m. Hence, we detected the response of an ensemble of NCs and not the dynamics of an individual crystal. The time-resolved XRD measurement is sketched in Figure~\ref{fig:NLelasticity}~d). Note that the diffracted beam is smeared out perpendicular to the diffraction plane on the detector. To quantify the smearing we transform the raw detector image in Figure~\ref{fig:NLelasticity}~d) into reciprocal space. Figure~\ref{fig:NLelasticity}~e) depicts the RSM of the NC streak in the q$_y$ and q$_z$ direction. The mosaicity broadening is attributed to a misalignment of the individual NCs [c.f. inset of Figure~\ref{fig:NLelasticity}~d)] that was already observed elsewhere for InP NCs on Si tips~\cite{Niu2016,Niu2019}. The broadening is also observed for different material systems, such as GaP~\cite{Kafi2024} and GaAs~\cite{Rodrigues2025} NCs on Si nanotip substrates. Furthermore, the streak is modulated by bright spots, which represent diffraction from either individual NCs or a sub-ensemble of NCs with better relative alignment~\cite{Niu2019}. In the following, we refer to these spots as hotspots and analyze their time-dependent intensity change [c.f. Figure S2].

\begin{figure*}[htbp]
    \centering
	\includegraphics[width = \textwidth]{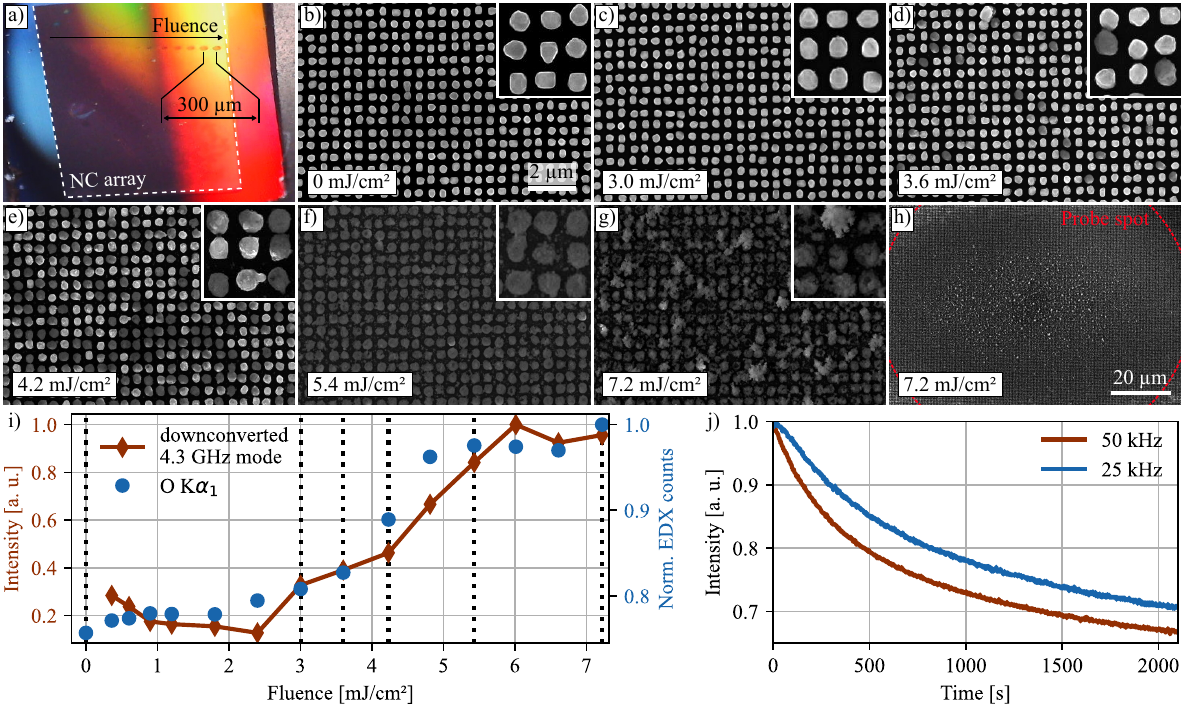}
	\caption{\textbf{(a)} Macroscopic image of \textit{Sample 3} taken after a series of measurements with 14 different fluences between 0.3\,mJ/cm$^{2}$ and 7.2\,mJ/cm$^{2}$. The perimeter of the NC array on the substrate is marked by a white dashed line. The fluence is increased for every spot on the sample with a distance from spot to spot of 300\,$\mu$m. \textbf{(b)} - \textbf{(g)} Close up SEM top-view images of the illumination area after exposure with different fluences. Structural modifications set in at a fluence of 3.6\,mJ/cm$^{2}$. After exposure to 5.4\,mJ/cm$^{2}$ all NCs in the image show structural degradation. \textbf{(h)} Overview SEM image of the spot with the highest fluence shows a localized area of damaged NCs due to the Gaussian intensity distribution of the pump laser spot. The probed area is sketched as red dashed ellipse. \textbf{(i)} EDX measurements of the oxygen K$\alpha_1$ emission line reveal a correlation of the NC oxidation and the intensity of the nonlinear downconverted mode in the photoacoustic signal. Vertical dashed lines mark the fluences that are shown in the SEM images \textbf{(b)} - \textbf{(g)}. \textbf{(j)} Real time degradation of the InP NCs in the laser beam over time. The laser was set to a repetition rate of 50\,kHz and 25\,kHz at a constant fluence of 0.5\,mJ/cm$^{2}$.}
	\label{fig:FluenceDep}
\end{figure*}

In the XRD measurement, we select a high excitation fluence of 7\,mJ/cm$^{2}$ to ensure excitation above the threshold for nonlinear frequency conversion. Figure~\ref{fig:NLelasticity}~f) depicts the Fourier-transformed time-dependent photoacoustic response for different hotspots in the detector image. All traces display a strong signal at a frequency around 0.7\,GHz. This frequency corresponds to the low-frequency mode already observed in the all-optical measurements [c.f.~\ref{fig:NLelasticity}~b) and c)]. Note that this mode has the highest amplitude of all non-fundamental modes.

While we can easily resolve the low-frequency response of the NC vibration, the frequency resolution is now limited to frequencies below 10\,GHz due to the duration of the 100\,ps X-ray probe pulse. To circumvent this limitation, we installed a PicoSwitch pulse slicer (TXproducts) to reduce the duration of the probe pulse to only 10\,ps~\cite{Sand2019a}. The inset in Figure~\ref{fig:NLelasticity}~f) depicts the experimental result with the shortened probe pulse. The corresponding transient change in intensity is shown in Figure S3 in the supplementary information. We observe a frequency component centered at 9\,GHz, i.e. near the frequency of the fundamental eigenmodes of the NCs of \textit{Sample 2}. Due to limited beamtime available for this measurement, we had to restrict the delay range to 200\,ps which comprises only two oscillation periods of the high frequency fundamental mode of \textit{Sample 2}. Thus, the frequency resolution of the measurement is insufficient to resolve both fundamental modes separately.

In addition to the transient intensity changes of individual hotspots we analyzed their transient position in the out-of-plane direction q$_z$ by integrating the 3D RSM along the in-plane directions q$_x$ and q$_y$. Exemplary results of the transient shift of the Bragg reflex measured on hotspot 2 are depicted in Figure S4 in the supplementary information in time- and frequency-domain.  

To summarize this section, we observe the appearance of new spectral components in the pump-probe measurements at high excitation fluence. Remarkably, we observe a low-frequency mode reminiscent of a difference frequency generation, e.g. in nonlinear optics. Our synchrotron-based time-resolved XRD measurements confirm the InP NCs as the origin of this mode, which we observe in the transient intensity as well as in a shift of the transient intensity in q$_z$-direction of the reciprocal space.

\begin{figure*}[t]
	\centering
	\includegraphics[width = \textwidth]{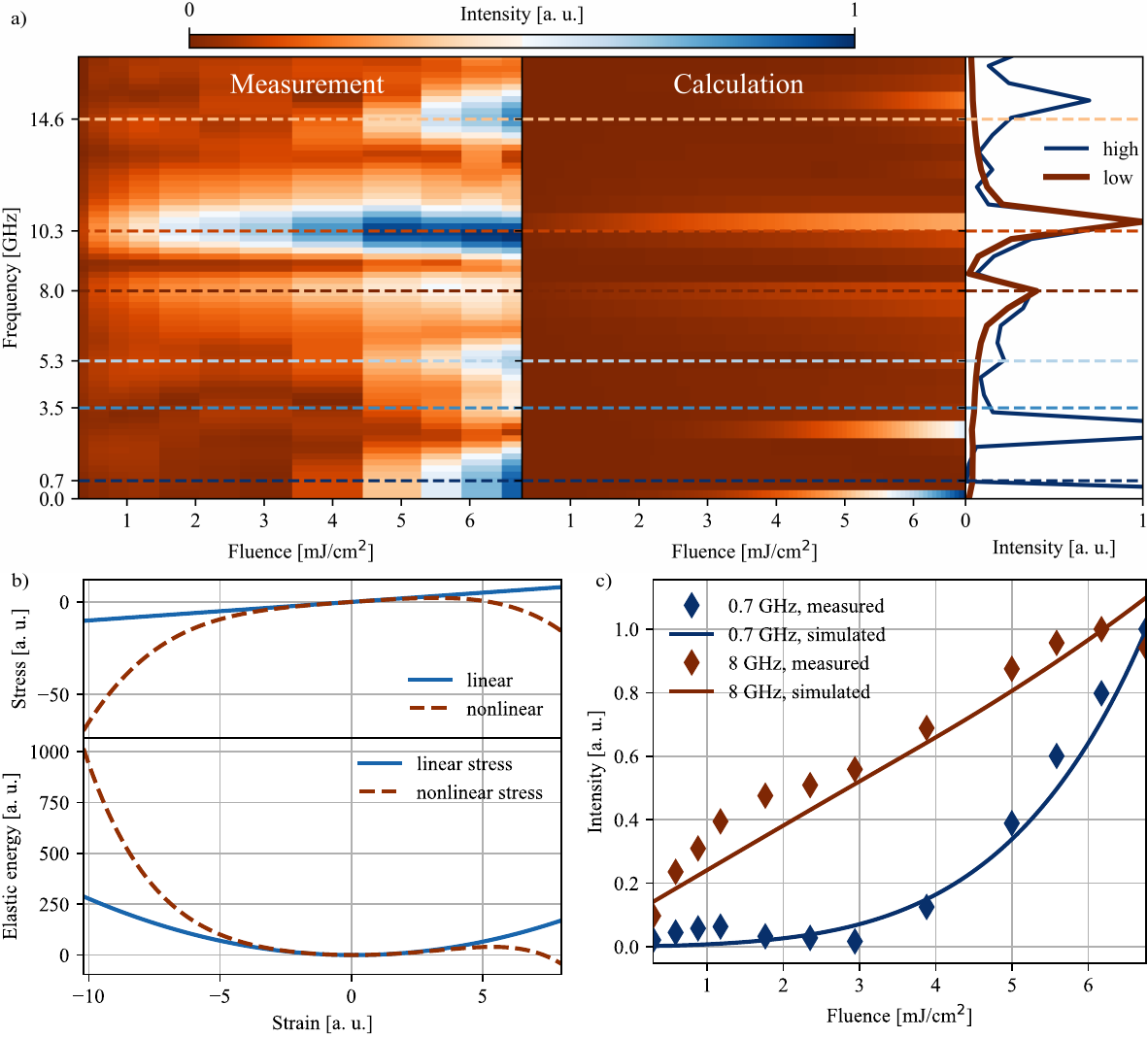}
	\caption{\textbf{(a)} Comparison of measured (left) and calculated (middle) fluence dependent photoacoustic response from the NCs in the spectral domain. The right panel displays the measured frequency spectra for high (blue) and low (red) excitation fluence. The simulation assumes a sinusoidal shape of the high-frequency (10.3\,GHz) and low-frequency (8\,GHz) fundamental breathing mode. Additional frequency components stem from nonlinear mixing of the excited acoustic modes. \textbf{(b)} Calculated nonlinear stress-strain relation (upper) and elastic energy (lower) using equ. (2) with the nonlinear coefficients $D_{1}=-0.05$, $D_{2}=0$ and $D_{3}=-0.005$. \textbf{(c)} Measured (diamonds) and calculated (solid line) fluence-dependent intensity of the fundamental low-frequency (red) and downconverted (blue) non-fundamental mode.}
    \label{fig:NonlinearElasticity}
\end{figure*}

\subsection{High Optical Fluence and Structural Modifications}\label{sec:Fluence}
In the following paragraphs, we will discuss the accumulated structural modification and its correlation to the non-fundamental modes in our time-resolved experiments with high excitation fluence. For that, we have investigated the structure ex-situ on \textit{Sample 3} after exposure to different excitation fluences. Specifically, we have exposed the sample to the excitation laser for 300\,s, i.e. the typical duration of a delay scan of the all-optical measurement. The laser was set to a low excitation fluence of 0.3\,mJ/cm$^{2}$. Next, we translate the sample by 300\,$\mu$m and expose it again to the excitation laser, now at a fluence of 0.6\,mJ/cm$^{2}$. We have repeated these steps with an exposure time of 300\,s until we reached a fluence of 7.2\,mJ/cm$^{2}$. Figure~\ref{fig:FluenceDep}~a) depicts a macroscopic view of the sample with the laser illumination marks on the surface. These marks stem from illumination with high laser fluence. Figure~\ref{fig:FluenceDep}~b)-g) depict top-view SEM images recorded ex-situ at specific illumination points. A laser fluence below 3\,mJ/cm$^{2}$ does not alter the sample structure visibly in the SEM image. We observe a noticeable change in shape and contrast of some NCs at 3.6\,mJ/cm$^{2}$. At 4.2\,mJ/cm$^{2}$ the contrast of more NCs has changed and the surface appears rougher. The overall shape of the NCs is still preserved at this fluence. The contrast of all NCs in the image frame has changed at 5.4\,mJ/cm$^{2}$. Some NCs were stressed above their elastic threshold and show a permanent deformation. Parts of crystals are torn off and lay in the voids of the NC array. The shape of the NCs is destroyed at the highest fluence of 7.2\,mJ/cm$^{2}$. Remarkably, we still observe intensity of the fundamental breathing modes, which are at 10.6\,GHz and 15\,GHz for \textit{Sample 3} due to the smaller NC dimensions. The Gaussian intensity distribution of the pump laser spot leads to an inhomogeneous excitation of the ensemble of NCs within the probe spot. Figure~\ref{fig:FluenceDep}~h) depicts the illumination spot with the highest fluence on a larger length scale. To get a sense of the illuminated area we sketched the outline of the probe spot with a size of 105\,$\mu$m x 98\,$\mu$m (red dashed ellipse). The pump spot covers more than three times of this area. We observe a localized damage of the NCs in the center of the image, which means that intact NCs further away from the center can still contribute to the observed transient signal. It is important to note that we refer to the beam's full-width-at-half-maximum (FWHM) size to calculate the excitation fluence. Thus, the local fluence in the area showing NC destruction is significantly higher. To investigate the origin of the damage mechanism we performed one-dimensional heat-diffusion simulations on a model structure that mimics \textit{Sample 3}. Optical excitation with a short laser pulse generates a temperature jump of 25.7\,K/(mJ/cm$^{2})$. Thus, for the highest fluence of 7.2\,mJ/cm$^{2}$ the initial temperature increase amount to $\Delta$T=185\,K. At a laser  repetition rate of 50\,kHz we observe no noticeable permanent heat of the NCs~\cite{Shayduk2020}. These results point towards laser ablation as a possible damage mechanism rather than melting of the NCs.

We performed complementary EDX measurements to investigate the elemental composition of the sample at the illumination points. We discover a correlation of the nonlinear response with the oxidation state of the NC. Figure~\ref{fig:FluenceDep}~i) depicts the oxygen (O) content (blue dots) obtained from the EDX measurement at different excitation fluences. Above 2\,mJ/cm$^{2}$ the O content increases and becomes very pronounced above 4\,mJ/cm$^{2}$, probably because of an increased surface area. We compare the fluence dependence of the O content with the strength of the downconverted mechanical mode (red solid line) that we measured in the photoacoustics measurement similar to \textit{Sample 1} in Figure~\ref{fig:NLelasticity}~c). Both curves show a qualitatively similar fluence dependence. Note that the NC size on \textit{Sample 3} is smaller which leads to higher frequencies of the fundamental breathing modes. Hence, the non-fundamental mode that corresponds to the difference frequency of the fundamental modes appears at a higher frequency of 4.3\,GHz.

As mentioned earlier, the temperature increase after the optical excitation prefers a photon induced process as the origin of the oxidation instead of the temperature~\cite{Wager1980}. Note that even a fluence of 0.4\,mJ/cm$^{2}$ leads to an increase in the oxygen level compared to an unexposed spot in Figure~\ref{fig:FluenceDep}~i). In fact, the top atomic layer of the InP NC does reach a sufficiently high temperature for thermal oxidation at an excitation fluence of 7.2\,mJ/cm$^{2}$ [c.f.~Figure S5]. However, this elevated temperature is maintained for less than 1\,ns before thermal equilibration reduces the average temperature below the oxidation threshold. To disentangle the influence of NC oxidation the appearance of non-fundamental acoustic frequencies modes of the NCs further experiments, e.g. in vacuum conditions, must be conducted. 

We also measured the real-time decay of the transient optical reflectivity at a positive pump-probe delay at two different laser repetition rates. The data is depicted in Figure~\ref{fig:FluenceDep}~j). The measurement was performed on \textit{Sample 1} at a low excitation fluence of 0.5\,mJ/cm$^{2}$ and at laser repetition rates of 50\,kHz and 25\,kHz, respectively. In both measurements, the average heating of the sample in the laser beam remains negligible. Note that the decay rate increases with the repetition rate, which indicates a photon flux dependent accumulation effect. The formation of a permanent oxide layer prevents the reflected intensity in all experiments to recover to its initial value. The intensity saturation of the fundamental modes as observed in Figure~\ref{fig:NLelasticity}~c) can be explained by the combination of two effects with increasing fluence. First, the number of NCs within the probe spot that can contribute to the transient signal is reduced as depicted in Figure~\ref{fig:FluenceDep}~h). Second, according to Figure~\ref{fig:FluenceDep}~j) the real-time decay of the reflected intensity is accelerated with increasing power input. Note that we have repeated the time-resolved optical measurements in different variants, e.g. reversing the order of the laser fluence and selecting unexposed areas on the sample for each measurement of a fluence series. The previously described picosecond dynamics persists in all measurements. 

\subsection{Nonlinear Elasticity Model}\label{sec:NLelasticity}

We now discuss the nonlinear frequency conversion of the acoustic modes at high optical excitation fluences. For that we quickly summarize the main features of the fluence dependence observed in the measurement and depicted in Figure~\ref{fig:NLelasticity}~c). The intensity of the two fundamental modes at 8\,GHz and 10.3\,GHz, respectively, increases linearly with fluence until it saturates at about 5\,mJ/cm$^{2}$. Above a fluence of 3\,mJ/cm$^{2}$ we observe non-fundamental modes with frequencies above and below the fundamental modes which increase super-linearly. Such nonlinear generation of frequency components is a well-known phenomenon in optics~\cite{Boyd2020} and similar effects were already observed for propagating sound~\cite{boja2015a}.

In the following we apply the well-established formalism of a nonlinear optical susceptibility to the propagation of an acoustic pulse. We begin by introducing the linear stress-strain relation which is also known as Hooke's law~\cite{Gros2018a}
\begin{equation}
    \sigma = E\cdot\eta,
    \label{equ:Hook_lin}
\end{equation}
where $\sigma$ and $E$ denote the mechanical stress and Young's modulus, respectively. $\eta = \frac{\Delta x}{x_{0}}$ denotes the deformation (strain). To allow for nonlinear frequency conversion of the acoustic wave we expand Hooke's law with nonlinear terms
\begin{equation}
    \sigma=E\cdot\eta+\sum_{n>0}D_{n}\cdot\eta^{n+1},
    \label{equ:Hook_nonlin}
\end{equation}
where $D_{n}$ denote the nonlinear coefficients. We employ Equ.~\ref{equ:Hook_nonlin} to calculate the nonlinear frequency response from the InP NCs. We employ harmonic signals at the frequency of the fundamental breathing modes as time-dependent strain. Figure~\ref{fig:NonlinearElasticity}~a) depicts the measured (left side) and calculated (middle) spectral response of the NCs. The panel on the right depicts the spectral response at high (blue) and low (red) excitation fluence. For better comparability we normalized both spectra to the maximum intensity of the high-frequency fundamental mode at 10.3\,GHz. Note that the fundamental modes of the NCs are independent of the excitation fluence. As explained in section~\ref{sec:Fluence} we observe a non-vanishing intensity of the fundamental modes even at high fluences despite the destruction of individual NCs. The calculated mixing products agree quantitatively with the experiment with the exception of the spectral width and intensity of the frequency components. The deviation stems from the harmonic input signal assumed in the calculation.

We used the experimental data to determine the nonlinear coefficients $D_{n}$ in Equ.~\ref{equ:Hook_nonlin}. As an example, we depict the measured and calculated fluence dependence of the low-frequency fundamental mode and the nonlinear difference frequency signal in Figure~\ref{fig:NonlinearElasticity}~c). The measurement was performed on \textit{Sample 1}. We find values $D_{1}=-0.05$ and $D_{3}=-0.005$ for the nonlinear coefficients in equation~\ref{equ:Hook_nonlin}. The resulting stress-strain relation and the elastic energy potential are depicted in Figure~\ref{fig:NonlinearElasticity}~b) in the blue (linear) and red (nonlinear) line. The elastic potential is calculated by integrating equ.~\ref{equ:Hook_nonlin} along an expansion $\Delta x$. The negative nonlinear coefficients lead to an inversion symmetry of the elastic potential with a repulsive force upon compression and an attractive force upon expansion. The potential diverges for large expansive strains beyond the deformation exerted in the experiment.

We want to stress phenomenological character of this model. We don't expect it to reproduce the complex broadband acoustic response of the NCs after femtosecond laser excitation, nor to determine quantitative nonlinear elastic coefficients of InP. For that a more advanced modeling beyond the scope of this work is necessary. Instead, the model performs a sanity-check of the fluence dependence of the non-fundamental modes and validates the symmetry assumptions of the underlying elastic potential. Both aspects point towards a possible nonlinear acoustic response which merits further investigation.

\section{Conclusion}\label{sec:Conclusion}
Photoacoustic excitation of InP NCs reveals nonlinear elastic behavior at high optical fluences, characterized by frequency mixing and the emergence of low-frequency acoustic modes. A nonlinear extension of Hooke's law successfully models this behavior and serves as a first step towards more sophisticated models that consider InP properties to determine the interatomic potential. Time-resolved XRD confirms the nanocrystals as the origin of the nonlinear response and supports acoustic decoupling from the Si nanotip substrate. The correlation between structural and elemental modifications and nonlinear mode generation, as observed via SEM and EDX, highlights the onset of damage mechanisms. These findings establish nonlinear phonon scattering as a sensitive probe of mechanical limits and material integrity in semiconductor nanostructures, offering new opportunities for nanoscale optomechanical control and advanced material diagnostics.

\section{Experimental Methods}\label{sec:Exp&Meth}
\begin{figure*}[t]
	\centering
	\includegraphics[width = \textwidth]{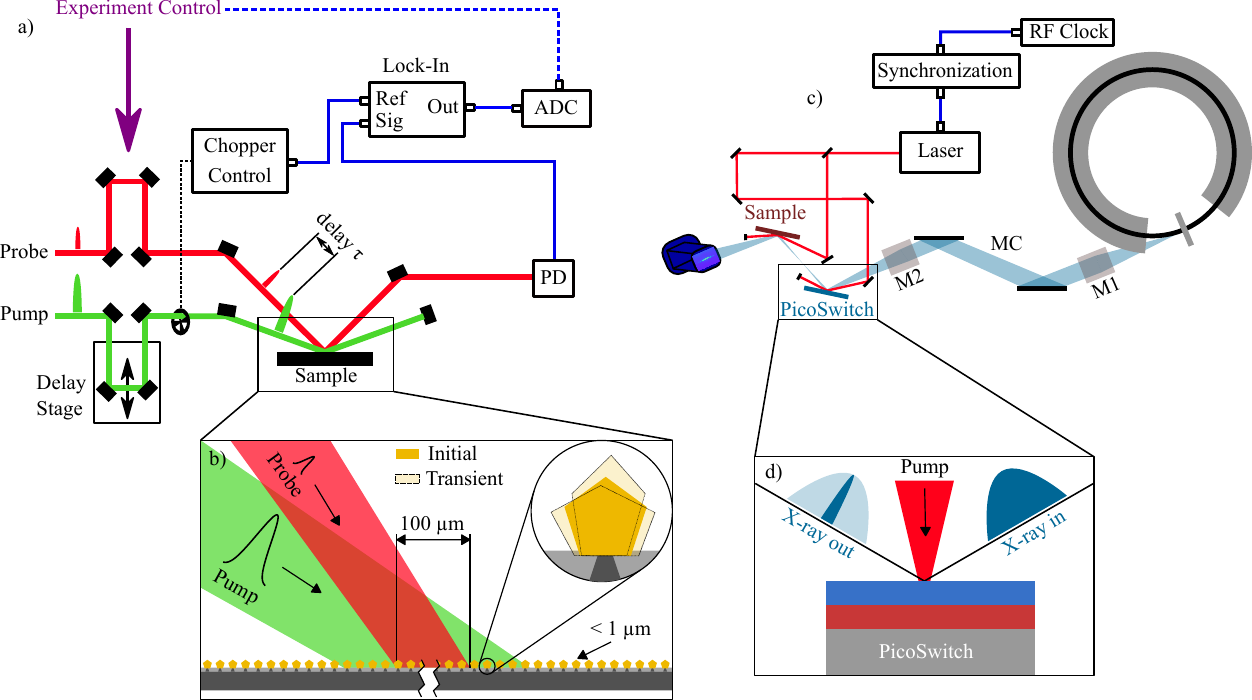}
	\caption{\textbf{(a)} Overview of the all-optical pump-probe setup. The pump and probe pulses are replica from the same laser source. The fundamental wavelength of the pump beam is 1030\,nm and it is frequency doubled in a BBO crystal to 515\,nm. The pump-probe delay is tuned with a motorized mechanical stage allowing a total delay range of 1\,ns. The pump beam is modulated with a mechanical chopper with a frequency of 1\,kHz. The chopper reference is fed to a lock-in amplifier and compared to the electronic output of the photodiode capturing the reflected probe pulse. \textbf{(b)} Detailed view of the sample with impinging pump and probe beams. The footprint of both beams is significantly larger than the size and pitch of the NCs. \textbf{(c)} Time-resolved XRD setup at ID09 beamline at ESRF. Details are given in the main text. \textbf{(d)} Slicing of the hard X-ray probe pulse using a PicoSwitch (TXproducts UG).}
	\label{fig:Experiment}
\end{figure*}

\begin{figure}[htbp]
	\centering
	\includegraphics[width = \columnwidth]{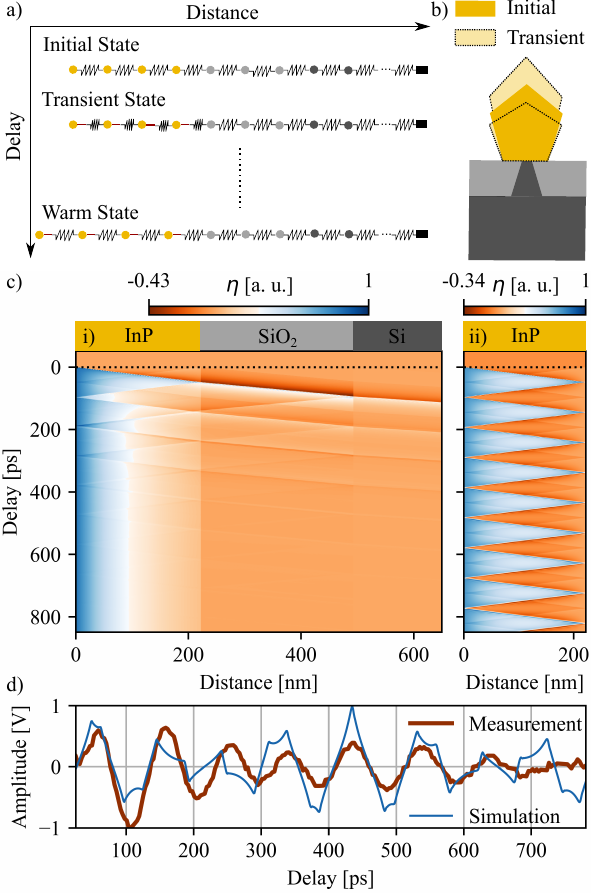}
	\caption{\textbf{(a)} Linear chain model of masses and springs to simulate the linear photoacoustic response of the NC. The first row (initial state) depicts the sample in equilibrium, the second row (transient state) depicts the displacive excitation and the third row (warm state) shows the sample in the displaced equilibrium. \textbf{(b)} Schematic representation of a NC in the initial (dark yellow) and in the transient (light yellow) state in one dimension. \textbf{(c)} Transient expansion (blue) and compression (red) along the linear chain of masses and springs (horizontal axis) for different pump-probe delay (vertical axis). We compared the response of different acoustic coupling of the NC to the substrate: i) The InP NC is coupled to the SiO$_2$ spacing layer and to the substrate. The acoustic energy dissipates into the depth of the substrate ii) The InP NC is fully decoupled. The acoustic energy remains confined in the nanostructure. \textbf{(d)} Comparison of the measured (dark red) and simulated (blue) linear photoacoustic response. The simulation is a linear superposition of the calculated low-frequency and high-frequency breathing modes.}
	\label{fig:MassAndSpringModel}
\end{figure}

\textbf{Materials:}
InP nanocrystals (NCs) were selectively grown on nanotip-patterned Si (001) wafers by gas-source molecular beam epitaxy via nanoheteroepitaxy, using solid indium and thermally cracked phosphine (PH$_3$). The Si nanotips, embedded in SiO$_2$, were arranged in square arrays with tip-to-tip distances ranging from 500\,nm to 2\,$\mu$m. Further details on wafer fabrication and epitaxy are provided in Refs.~\cite{Niu2016,Skibitzki2017,Niu2019}. The surface morphology was examined using a SEM system operated at 10 kV (Pioneer Two System, Raith Fabrication). The NC dimensions were determined from SEM images using the image processing and analysis software ImageJ. From top-view images we measured the NC area and calculated an equivalent diameter by assuming a circular shape of the crystals. The height of the NCs was measured from the lowest to the highest point in cross-section SEM images. The equivalent diameter and height of an average NC for all samples used in this work are listed in table~\ref{tab:SampleParameters}.

The chemical composition of the surface was analyzed by EDX using a GeminiSEM 500 microscope (Zeiss) equipped with an InLens detector and operated at a beam energy of 10 keV. The elemental analysis was carried out with a spectrometer (XFlash 6130, Bruker).

\textbf{Breathing modes in particles with spherical shape:}
The shape of a NC can be approximated as nearly spherical. The eigenfrequencies of an elastic sphere can be calculated analytically~\cite{Lamb1881}. The frequency of a radial mode in the continuum approximation is given by $f=\tau_n\nu_L/(\pi D)$ where $\tau_n$ are the eigenvalues of the equation $\tau_ncot(\tau_n)=1-\nu_L^2\tau_n^2/(4\nu_T^2)$. $\nu_L$ and $\nu_T$ are the longitudinal and transverse sound velocities and $D$ is the diameter of the sphere. An ideal sphere has only one fundamental breathing mode and additional higher order vibrations. Our NCs show a height-to-diameter ratio < 1 [see table~\ref{tab:SampleParameters}]. Therefore we expect the appearance of two modes from the calculation similar to the measurement shown in Figure~\ref{fig:PAresponse}~c). With material parameters for InP~\cite{NICHOLS1980667} we calculate $\tau_1$ as 2.241 for the fundamental radial breathing mode. Adding the height and diameter of an average NC of each sample yields the predicted frequency of both modes listed in table~\ref{tab:SampleParameters}. The calculated frequencies are typically lower then the measured frequencies. This can be in part explained by the overestimated diameter and height of the NCs due to the precision of the measurement of the NC size as explained above. This is especially true for \textit{Sample 1} because of the occurrence of connected or overlapping NCs. \textit{Sample 2} shows a higher calculated frequency for both fundamental modes. \textit{Sample 1} and \textit{Sample 2} originate from the same fabrication run, with the pitch size being the only parameter that differs. It has been observed previously that that an increased pitch size leads to slightly reduced dimensions of the NCs in these kind of samples~\cite{Kafi2024}. Therefore, we expect higher frequencies of the fundamental modes for \textit{Sample 2}. The SEM images and the pump-probe measurement were not conducted on the same spot on the samples. The deviation between the measured and calculated frequencies can then only be explained by an inhomogeneous NC size distribution across the NC array.

\textbf{Pump-Probe Measurements:}
To excite vibrational modes of the NCs, we illuminate the sample with femtosecond light pulses from a frequency doubled femtosecond Yb:YAG amplified laser (Pharos, Light Conversion). The fundamental and second harmonic wavelength is 1030\,nm and 515\,nm, respectively. We monitor the transient sample response by time-resolved optical pump-probe reflectivity. The experiment was conducted at a repetition rate of 50\,kHz. The experimental setup is depicted in Figure~\ref{fig:Experiment}~a) and b). Due to the probe footprint of 111\,$\mu$m x 104\,$\mu$m on \textit{Sample 1} we measure a transient deformation of an ensemble of NCs on the order of 10$^{4}$. The footprint of the pump laser on \textit{Sample 1} was 238\,$\mu$m x 182\,$\mu$m and the average power of the pump laser was set between 5\,mW and 115\,mW. A similar spot size and average power was used on \textit{Sample 3}. We employed a mechanical translation stage (UTS150CC, Newport) to tune the pump-probe delay. The travel range of 150\,mm is equivalent to a maximum delay of 1\,ns and we set the accessible positive pump-probe delay to 850 ps. The second harmonic of the pump beam was generated by a beta barium borate (BBO) crystal (NLC06, Thorlabs). The temporal resolution of the setup was limited by the laser pulse length of about 270\,fs. We detected changes in the transient reflectivity of the NCs with a biased Si photodiode (DET10A2, Thorlabs). For a better signal-to-noise ratio we modulated the pump beam with a mechanical chopper at a rate of 1\,kHz and compared this reference signal to the electronic output of the photodiode in a lock-in amplifier (LIA-MV-150, Femto).

To collect complementary data on longer timescales, we employed optical pump - X-ray probe reflectivity measurements at the ID09 time-resolved beamline (ESRF). In the standard configuration of the beamline the temporal resolution was limited by the X-ray probe pulse duration to about 100\,ps. The pump pulses were generated with a femtosecond laser (Legend, Coherent) with a central wavelength of 800\,nm. The laser was electronically phase locked to the storage ring and a mechanical chopper reduced the repetition rate of the experiment to 986\,Hz. We recorded the transient evolution of the symmetric (004) Bragg reflex with a CCD area detector (MX170-HS, Rayonix) at an energy of 15\,keV. This pump-probe scheme is conceptually equal to the all-optical setup, however, the X-ray probe beam is directly sensitive to the structural deformation by probing the transient lattice parameter of the NCs. To overcome the temporal resolution limit of the beamline and to resolve acoustic modes of the NCs with higher frequencies we used a fast Bragg switch (PicoSwitch, TXproducts UG). The experimental setup in the high-resolution configuration is shown in Figure~\ref{fig:Experiment}~c) and d). The X-ray pulse duration was shortened to less than 10\,ps. The PicoSwitch employs a photoacoustic excitation on a specifically designed crystalline heterostructure. Absorption of the laser pulse triggers coherent acoustic phonon wavepackets, similar to the excitation of acoustic modes in the InP NCs depicted in Figure~\ref{fig:PAresponse}~c). These propagating sound waves modulate the lattice parameter of a switching layer, thus shifting diffraction intensity in reciprocal space on a picosecond timescale. A comprehensive demonstration of the PicoSwitch device is published elsewhere~\cite{Sand2019a,Sand2016a}.

\textbf{Photoacoustic Simulations:}
In the following, we discuss model calculations of the picosecond structural dynamics. Specifically, we investigate the origin of the long-lived oscillations in the all-optical reflectivity data and in the time-resolved XRD experiment. The simulation of the coherent acoustic response employs a one-dimensional chain of coupled harmonic oscillators to model the sample structure~\cite{herz2012b} like it is shown in Figure~\ref{fig:MassAndSpringModel} a). We employ the udkm1Dsim python toolbox~\cite{Schi2021a} to perform our calculations on two different structural models. The simulation only considers one-dimensional phonon propagation perpendicular to the sample surface. Thus, we simulate the dynamics of the low-frequency and high-frequency breathing mode individually by assuming different film thicknesses in the calculation. Figure~\ref{fig:MassAndSpringModel} b) depicts the one-dimensional breathing of a NC assumed for our simulation. In a second step we compare the sum of both simulations to the experimental data. We perform this simulation for two different structures and calculate the average strain in the InP layer. First we model the NC as an InP layer on a Si substrate. The simulation shows a strong acoustic coupling between the NC and the substrate [c.f. Figure~\ref{fig:MassAndSpringModel}~c) panel i)]. Here, the coherent phonon wavepacket quickly dissipates from the film to the substrate. Second we model the sample as a stand-alone NC, i.e. only the InP film [c.f. Figure~\ref{fig:MassAndSpringModel}~c) panel ii)] so that no acoustic energy can dissipate into the substrate. In Figure~\ref{fig:MassAndSpringModel}~d) we compare the time-resolved reflectivity measurement to the simulation. The measured oscillations persist for more than 800\,ps, i.e. the full delay range of the measurement. We obtain a qualitatively similar response only from our second simulation case which suggests that the NC is acoustically decoupled from the substrate. Thus, any effect due to propagation of the acoustic wave in the structure must stem from interactions in the InP NC.
The simulation parameters are listed in table~\ref{tab:SimParameter}.

\begin{table}[h]
    \centering
    \resizebox{\columnwidth}{!}{
    \begin{tabular}{|c|c|c|c|c|}\hline
     
     \textbf{Parameter Name}&\textbf{Unit} & \textbf{InP}& \textbf{SiO$_{2}$} &\textbf{Si}\\\hline
     Lattice Parameter&\,nm&0.586945~\cite{https://doi.org/10.1002/pssa.2211030214}&\begin{tabular}{@{}c@{}}c = 0.54163~\cite{+1992+177+212} \\ a = 0.4921~\cite{+1992+177+212}\end{tabular}&0.5431~\cite{RevModPhys.59.1121}\\\hline
     Long. Sound Velocity [001]&\,nm/ps&4.6~\cite{NICHOLS1980667}&5.97~\cite{10.1063/1.1721449}&8.43~\cite{10.1063/1.1702809}\\\hline
     Refractive Index @ 515 nm&1&3.7506+0.46117i~\cite{PhysRevB.27.985}&1.4616~\cite{Arosa:20}&4.2231+0.061005i~\cite{PhysRevB.27.985}\\\hline
     Thermal Conductivity&\,W/(m\,K)&68~\cite{PhysRev.133.A1665}& 1.4~\cite{10.1063/1.112355}&156~\cite{PhysRev.134.A1058}\\\hline
     Thermal Expansion Coefficient&\,1/K&4.48E-6~\cite{https://doi.org/10.1002/pssa.2211030214}& 0.56E-6~\cite{ElKareh1994}&2.616E-6~\cite{10.1063/1.323747}\\\hline
     Heat Capacity&\,J/(kg\,K)&306~\cite{Piesbergen+1963+141+147}&725~\cite{Andersson_1992}&710~\cite{Flubacher01031959}\\\hline
         
    \end{tabular}}
\caption{Material parameters used in the photoacoustic simulation of a 1-dimensional sample structure along the [001] direction.}
\label{tab:SimParameter}

\end{table}

\bmsection*{Author contributions}
DH built the experimental setup and conducted the all-optical measurements. DH, DS, MB, and PG carried out the synchrotron measurements. AR, AK, OS, and FH grew the samples and performed nanolithography to assemble the nanocrystal arrays. AR, OS, and FH characterized the samples using SEM and static PL measurements. DH and PG performed photoacoustic simulations and modeled the nonlinear acoustic response. PG conceived the experiment and wrote the first draft of the manuscript. FH and PG supervised the project and acquired funding. All authors contributed to the writing of the manuscript and to the interpretation of the results.

\bmsection*{Acknowledgments}
We acknowledge the European Synchrotron Radiation Facility (ESRF) for provision of synchrotron radiation facilities under proposal number HC-5502, and we would like to thank Dr. Matteo Levantino and Dr. Celine Mariette for assistance and support in using beamline ID09. We thank Kristiane Elsner for the measurement of the elemental sample composition. We thank Dr. Martin Handwerg (IKZ) for his careful reading of our manuscript. This work was funded by the Deutsche Forschungsgemeinschaft (DFG, German Research Foundation) – project numbers GA 2558/7-1, GA 2558/9-1, 428250328 and 471105220. 

\bmsection*{Financial disclosure}

None reported.

\bmsection*{Conflict of interest}

The authors declare no potential conflict of interests.

\bibliography{bibcollection_updated}

\bmsection*{Supporting information}

Data is made available upon request.

\end{document}